\documentclass{emulateapj}
\usepackage{apjfonts}
\usepackage{lscape}

\slugcomment{To appear in the Astrophysical Journal Letters}

\shorttitle{Nearby halo white dwarf with $\mu=2.55\arcsec$ yr$^{-1}$}
\shortauthors{Lepine, Rich, \& Shara}

\begin{document}

\title{Discovery of a nearby halo white dwarf with proper
  motion $\mu=2.55\arcsec$ yr$^{-1}$}

\author{S\'ebastien L\'epine\altaffilmark{1,2}, R. Michael
  Rich\altaffilmark{3}, and Michael M. Shara\altaffilmark{1}}

\altaffiltext{1}{Department of Astrophysics, Division of Physical
Sciences, American Museum of Natural History, Central Park West at
79th Street, New York, NY 10024, USA, lepine@amnh.org, mshara@amnh.org}
\altaffiltext{2}{Visiting Astronomer, Kitt Peak National Observatory}
\altaffiltext{3}{Department of Physics and Astronomy, University of
California at Los Angeles, Los Angeles, CA 90095, USA,
rmr@astro.ucla.edu}

\begin{abstract}
We report the discovery of PM J13420-3415, a faint (V=17) white dwarf
with a very high proper motion $\mu=2.55\arcsec$ yr$^{-1}$. The star
was found in the southern sky extension of the SUPERBLINK proper
motion survey. A red spectrum shows the classical signature of a DA
white dwarf, with a weak $H\alpha$ line in absorption as the only
prominent feature. The star is also found to have a large radial
velocity $V_{rad}=+212\pm15$km $s^{-1}$. At the adopted distance of
$18$pc, the star has a very large space motion of $313$ km s$^{-1}$
relative to the Sun. An integration of the space motion shows that
the star is on a nearly polar Galactic orbit, and is thus an
unambiguous member of the Galactic halo. However, with an estimated
effective temperature $5,000K<T_{eff}<5,500K$, the white dwarf appears
to be much younger than expected for a denizen of the halo. The
apparent paradox can be explained if the white dwarf is the relatively
young ($\sim2$Gyr) remnant of a longer-lived ($10-14$Gyr) main
sequence star, in which case the object is predicted to be a low-mass
white dwarf with $M\approx0.45M_{\odot}$.
\end{abstract}


\section{Introduction}
Halo white dwarfs are expected to be found in the Solar Neighborhood
($d<25pc$) as faint stars with very large proper motions. While their
number density is predicted to be low ($\sim2\times10^{-5}$
pc$^{-3}$), the empirical determination of their density and
luminosity function has significant implications for the formation and
evolution of the Galaxy. Halo white dwarfs are a candidate for
baryonic dark matter \citep{IIBSG00}; they can also be used to
constrain the age of the galactic halo \citep{FBB_01}.

Unfortunately, the current census of halo white dwarfs in the
Neighborhood of the Sun is limited to a small number of candidate
objects, and is probably very much incomplete at this time. A sample of
only six candidate halo white dwarfs was originally assembled by
\citet{LDM89} out of the very high proper motion stars listed in the
LHS catalog of \citet{L79}. Few additional high velocity white dwarfs
were identified in subsequent proper motion surveys. A pair of cool
(4000K), common proper motion ($\mu=1.9\arcsec$ yr$^{-1}$) white
dwarfs was discovered by \citet{SSAII_02}, but their status as
halo objects remain uncertain. A group of 38 high proper motion white
dwarfs were identified in a survey of the south Galactic cap
\citep{O01}, but their halo membership is under debate, as reviewed 
in \citet{K03} and \citet{HL03}. The main objections are based on the
apparent young age of the group, more consistent with a disk
population \citep{B03}, and about proper motion selection effects,
which suggest that most or all the stars in the group are extreme
members of the Galactic thick disk \citep{SCLB05}.

An independent approach to finding halo white dwarfs is based on the
identification of very old, ultra-cool objects from multiband
photometry \citep{Getal_04}. It remains true, however, that any
population of white dwarfs can only be reliably associated with the
Galactic halo based on the kinematics of the group, which does
require proper assessment of their radial velocities, distances, and
proper motions.

In this Letter, we report the identification of a faint white dwarf
with a very high proper motion ($\mu=2.55\arcsec$ yr$^{-1}$) and large
radial velocity ($V_{rad}=+212$km s$^{-1}$). The highly probable halo
membership of this white dwarf is discussed in terms of its extreme
kinematics and debated with respect to its apparently young age.

\section{Proper Motion Discovery and photometry}

The high proper motion star PM J13420-3415 was discovered as part of
the SUPERBLINK survey for high proper motion objects in the Digitized
Sky Survey (DSS), recently extended to southern declinations
\citet{L05a}. A detailed description of the survey method is found in
\citet{LS05}. The star was found by the SUPERBLINK software after
analysis of scans of SERC-J (blue) and SERC-SR (red) photographic
survey plates, separated by 16 years. A search for additional
sightings of the star yielded positive identification in one SERC-I
plate (also in the DSS), and in a 2MASS survey image. Figure 1
displays the discovery field at four different epochs spanning almost
25 years. 


\begin{figure}
\epsscale{1.15}
\plotone{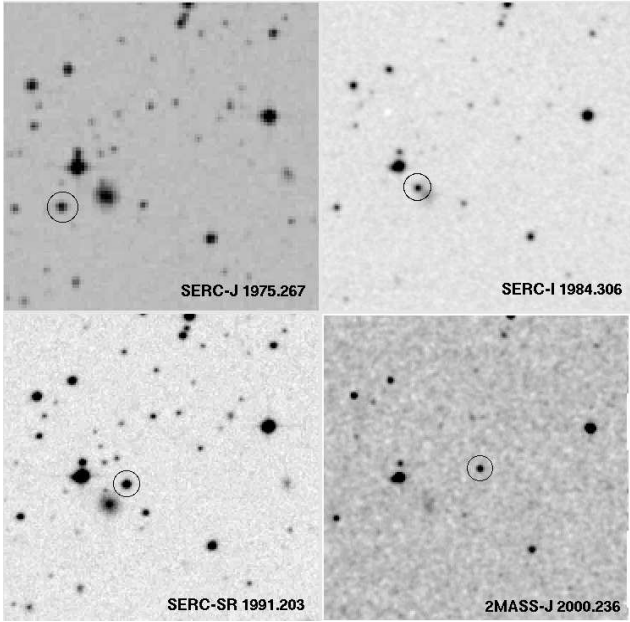}%
\caption{Four different sightings of the new high proper motion white
dwarf PM J13420-3415 ($\mu$=2.55$\arcsec$ yr$^{-1}$). The position of
the star is marked with a circle. Upper left: SERC-J photographic
survey, blue band (IIIaJ+GG395). Upper right SERC-I photographic
survey, near-infrared band (IVN+RG715). Lower left: SERC-SR
photographic survey, red band (IIIaF+GG395). Lower right: 2MASS
infrared CCD survey (J band image shown). The epoch of each survey
image is noted on the lower right. Each field is 3.15$\arcmin$ on the
side.}
\end{figure}

A proper motion of 2.55$\arcsec$ yr$^{-1}$ is calculated based on the
position of the star on the SERC-J DSS scan, used as the first epoch,
and its position recorded in the 2MASS All-Sky Catalog of Point
Sources \citep{Cetal03}, used as the second epoch. We derive a
transformation from the DSS scan XY positions into the 2MASS J2000
coordinate system by using a set of 5 moderately bright ($13<J<16$)
reference stars within 1.5$\arcmin$ of PM J13420-3415. The XY position
of PM J13420-3415 on the DSS scan was then transformed into its
corresponding J2000 coordinates in the 2MASS reference frame. We
calculate a proper motion of -2.275$\arcsec$ yr$^{-1}$ in the
direction of R.A., and 1.145$\arcsec$ yr$^{-1}$ in the direction of
Decl., both with an estimated error of 0.012$\arcsec$ yr$^{-1}$. Data
on the star are compiled in Table 1.

\begin{deluxetable}{lrl}
\tabletypesize{\scriptsize}
\tablecolumns{3} 
\tablewidth{0pt} 
\tablecaption{Basic Data for PM J13420-3415} 
\tablehead{Datum & Value & Units}
\startdata 
RA (2000.0) & 13 42  2.84 & h:m:s\\
DEC (2000.0)& -34 15 19.1 & d:m:s\\
$\mu_{RA}$       &-2.275 & $\arcsec$ yr$^{-1}$\\
$\mu_{Decl}$     & 1.145 & $\arcsec$ yr$^{-1}$\\
V$_{rad}$   & +212 $\pm$15 & km s$^{-1}$\\
B$_J$\tablenotemark{1}     &   17.5 $\pm$0.3  & mag\\
R$_F$\tablenotemark{1}     &   16.1 $\pm$0.3  & mag\\
J\tablenotemark{2}     &   15.00$\pm$0.02 & mag\\
H\tablenotemark{2}     &   14.75$\pm$0.02 & mag\\
K$_s$\tablenotemark{2} &   14.65$\pm$0.03 & mag\\ 
Spectral Type & DA9.5& \\
Distance & 18$_{-3}^{+7}$ & pc \\
$U$ &     +62 & km s$^{-1}$\\
$V$ &    -211 & km s$^{-1}$\\
$W$ &    +233 & km s$^{-1}$
\enddata
\tablenotetext{1}{Photographic blue ($IIIaJ$) and red ($IIIaF$)
  magnitudes from USNO B-1.0 catalog.}
\tablenotetext{2}{Infrared J, H, and K$_s$ magnitudes from 2MASS
  All-Sky Catalog of Point Sources.}
\end{deluxetable}

%
%
%
%

Photometry is obtained from the 2MASS All-Sky Catalog and from
the USNO-B1.0 catalog \citep{Metal03}. In USNO-B1.0, the star matches
the source 0557-0303180. Although this source is not listed in the
USNO-B1.0 as having any proper motion, 0557-0303180 clearly matches
the position of PM J13420-3415 at the epoch of the SERC-SR plate. The
source has a blue photographic magnitude $B_J=17.48$ and a red
photographic magnitude $R_F=16.12$. These are consistent with a visual
magnitude $V \simeq B_J - 0.46 ( B_J - R_F ) = 16.85$, with an
uncertainty of $0.3$ mag. In the 2MASS All-Sky Point Source Catalog,
PM J13420-3415 is a match to the point source 2MASS J13420283-3415190,
which has infrared magnitudes J=15.00, H=14.75, and K$_s$=14.65.

%
%
%

\begin{figure}
\epsscale{1.18}
\plotone{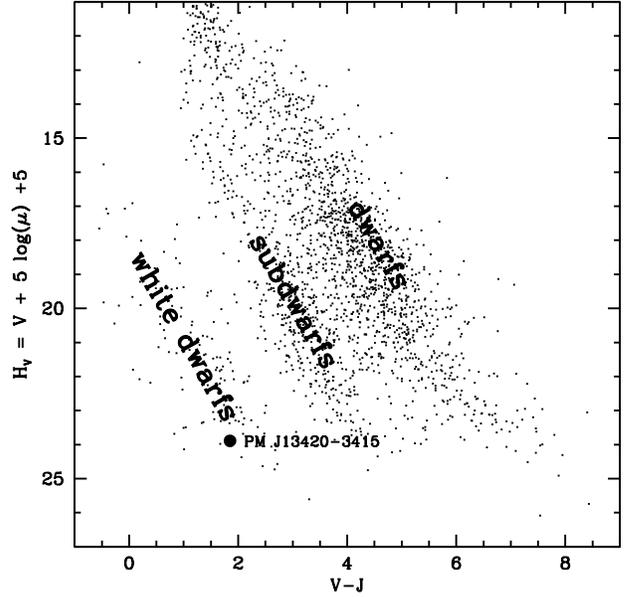}
\caption{Reduced proper motion diagram, with PM J13420-3415 superposed on
  the distribution of $\mu>0.5\arcsec$ yr$^{-1}$ stars from the
  LSPM-north catalog. PM J13420-3415 lies well into the locus of
  (cool) white dwarfs.}
\end{figure}

With a color index V-J=1.85, PM J13420-3415 is bluer than most faint
stars with very high proper motions. Figure 2 shows the position of PM
J13420-3415 in the $[H_V,V-J]$ reduced proper motion diagram. For
comparison, the diagram also shows the 2,194 stars in the LSPM-north
proper motion catalog \citep{LS05} that have $\mu>0.5\arcsec$
yr$^{-1}$. With its blue color and large proper motion, PM J13420-3415
clearly falls into the locus of the white dwarfs, amidst the cooler,
low-luminosity bodies.

\section{Spectroscopy, distance, and kinematics}

Spectroscopy was performed on the night of 26 June 2005 with the 4m
Mayall telescope on Kitt Peak. A spectrum of PM J13420-3415 was
obtained with the RC Spectrograph equipped with the LB1A thick CCD. We
used the 316 l/mm grating blazed at $7500$\AA, with an GG550 order
blocking filter, and a slit of 1.5$\arcsec$ giving a spectral
resolution of $1.74$\AA pixel$^{-1}$. Standard spectral reduction was
performed with IRAF using the CCDPROC and SPECRED packages, including
removal of telluric features. The spectrophotometric calibration is
based on observations of four standards from \citet{MG90}. The
target and standards were all observed with a slit angle within 15
degrees of the parallactic angle, so as to minimize slit loss due to
atmospheric diffraction.

%
%
%

\begin{figure}
\epsscale{1.18}
\plotone{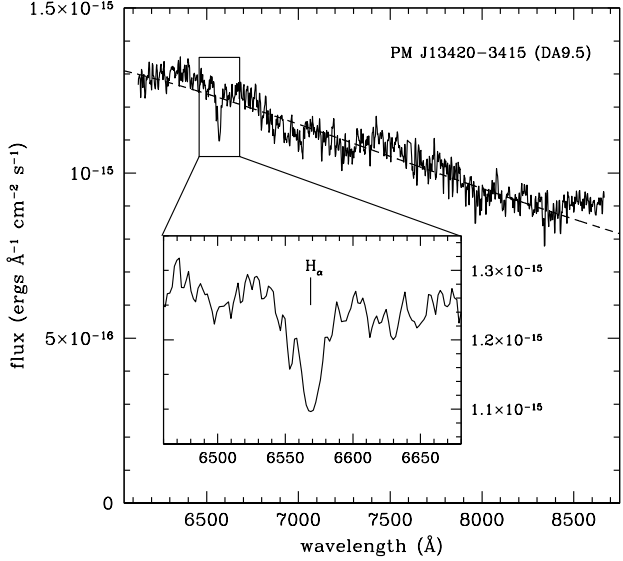}
\caption{Red optical spectrum of PM J13420-3415, with a clear
  detection of $H\alpha$ in absorption. The spectral energy
  distribution matches that of a $T_{eff}$=5500K blackbody (dashed
  curve). A radial velocity of +212$\pm$15 km s$^{-1}$ is derived
  from the shift in the $H\alpha$ line.}
\end{figure} 

The spectrum bears the signature of a cool, hydrogen white dwarf
(Figure 3). The only prominent feature is a shallow $H\alpha$
absorption line. The line is quite broad, with a full width at half
maximum of $21\pm2$\AA, and an equivalent width of $2.9$\AA. The
spectral energy distribution appears rugged at the $5\%$ level, with
broad bumps and troughs: these are possibly residuals from the removal
of sky lines and/or correction for telluric absorption features. The
star was observed at an airmass of $\approx2.6$, and both sky lines
and telluric absorption in that wavelength range were significant,
with corrections susceptible to large errors. 

We classify PM J13420-3415 as a hydrogen white dwarf with spectral
subtype DA9.5. A blackbody fit to the spectrum best matches an
effective temperature $T_{eff}=5500K$. The weakness of $H\alpha$
is however consistent with a slightly cooler $T_{eff}$. A comparison
with the models and observations presented in \citet{BRL_97} shows the
$H\alpha$ profile to be comparable to those of DA white dwarfs with
$T_{eff}$ in the 5200K-5500K range. This temperature scale is
corroborated by the photometry, based on a comparison with the
atmospheric models of \citet{BWB_95}. A derivation of CIT color
indices from 2MASS $JHK_s$ magnitudes\footnote{See the 2MASS
  documentation at http://www.ipac.caltech.edu/2mass/} yield
$V-K_{CIT}=2.18$, $J_{CIT}-H_{CIT}=0.274$, and $H_{CIT}-K_{CIT}=0.062$
for PM J13420-3415. With all weights equal, these values are most
consistent with the $T_{eff}=5000K$ atmospheric model
($V-K_{CIT}$=1.828, $J_{CIT}-H_{CIT}$=0.243,
$H_{CIT}-K_{CIT}$=0.056). However, our $V$ magnitude is derived from
photographic data and is much more uncertain than the 2MASS infrared
magnitudes. With more weight on $J_{CIT}-H_{CIT}$ and
$H_{CIT}-K_{CIT}$, PM J13420-3415 is more consistent with the
$T_{eff}=5500K$ model ($V-K_{CIT}$=1.650, $J_{CIT}-H_{CIT}$=0.260,
$H_{CIT}-K_{CIT}$=0.082). It is thus reasonable at this time to adopt
an intermediate value for $T_{eff}$, in line with a spectral subtype
DA9.5.

%
%
%
%
%
%
%

A radial velocity of +212$\pm$15 km s$^{-1}$ is calculated from
the observed redshift in the centroid of the H$\alpha$ absorption
line. A fit to the line profile is made using {\it splot} in IRAF. The
radial velocity is calibrated with a spectrum of the sdF8 standard BD
+17 4708, whose radial velocity is known to $\pm0.5$ km s$^{-1}$
\citep{LSTDMCLM_02}, and which we observed on the same night as PM
J13420-3415 and with the same instrumental setup. The radial velocity
for PM J13420-3415 includes a correction for the expected
gravitational redshift. While the surface gravity of the object is not
known precisely at this point, \citet{R96} finds gravitational
redshifts from field white dwarfs that are 28.3km s$^{-1}$ on average,
with a dispersion of only 3.9km s$^{-1}$; we thus adopt a -28.3km
s$^{-1}$ correction. A error $\pm15$ km s$^{-1}$ is estimated from the
quality of the centroid fit, the accuracy of the wavelength
calibration, and the dispersion in the range of possible gravitational
redshifts.


From the measured proper motion and radial velocity, the full
kinematics of PM J13420-3415 can be determined based on an estimate of
the distance. There exists several calibrations for absolute
magnitudes as a function of either color or $T_{eff}$. The
relationship defined by \citep{O01} uses photographic magnitudes $B_J$
and $R_F$ as follows: $M_{B_J}=12.73+2.58(B_J-R_F)$. Under this
calibration, PM J13420-3415 has an absolute magnitude $M_{B_J}=16.24$,
and resides at a distance d=17.7pc. A calibration of the $[M_{V},V-J]$
relationship for white dwarfs with known trigonometric parallaxes
\citep{L05b} suggests $M_{V}=15.5$, which would be consistent with a
distance d=18.6pc. The atmospheric models of \citep{BWB_95} suggest
$M_{V}=14.62$ for $T_{eff}=5500$ hydrogen white dwarfs, and
$M_{V}=15.11$ for $T_{eff}=5000$. An intermediate value would place
the star at a distance of d=24.9pc. The latter distance scale would
translate into a huge transverse velocity $V_{T}=301$km s$^{-1}$. A
more conservative distance scale $d\approx18pc$, consistent with the
two photometric distance estimates above, would yield a more modest
$V_{T}=216$km s$^{-1}$.

%
%
%
%
%
%
%
%
%

Using the more conservative distance range, we calculate the UVW
components of the velocity, where U is the velocity in the direction
of the Galactic center, V is in the direction of Galactic rotation,
and W is towards the north Galactic pole. We find $U=+57$km s$^{-1}$,
$V=-207$km s$^{-1}$, and $W=+220$km s$^{-1}$. These correspond to a
total space motion of 307 km $s^{-1}$ relative to the Sun. At the
larger distance scale of 25pc, the total space motion would be 357 km
$s^{-1}$. Note that a more conservative distance estimate of
15pc still yields a total space motion of 287 km $s^{-1}$, largely
because of the high value of $V_{rad}$.

\begin{figure}
\epsscale{1.18}
\plotone{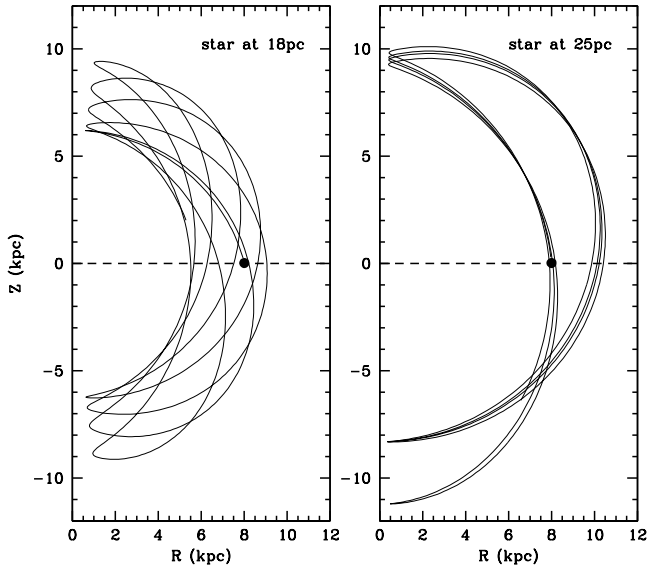}
\caption{Integrated Galactic orbital motion of PM J13420-3415, based
  on its estimated local kinematics (UVW velocity components). The
  orbit is displayed in the $[R,Z]$ plane, where $R$ is the
  galactocentric distance and $Z$ the distance from the
  midplane. The initial UVW are dependent on the estimated distance of
  the star from the Sun: two solutions and given for d=18pc (left),
  and d=25pc (right). In both cases (and for all $18pc<d<25pc$), the
  star evolves on a nearly polar orbit, and is thus clearly a denizen
  of the Galactic halo. Whether the star or its progenitor actually
  originated in the halo remains an open question.}
\end{figure} 

From the adopted UVW velocity components, we integrate the Galactic
orbital motion using the Galactic mass model of \citet{DC95}, which
includes separate terms for the bulge, disk, and halo. We use a
Runge-Kutta fourth order integrator in time steps of $10^3$
yr. Figure 4 shows 800 Myr integrations (for both the 18pc and 25pc
distance estimates) of the orbit of the star plotted in the [R,z]
plane, where R is the galactocentric distance in cylindrical
coordinates, and z is the distance from the plane. In both cases (and
at all intermediate distance ranges), the star is found to evolve on a
nearly circumpolar orbit, overwhelmingly consistent with a halo
membership.

\section{Discussion and conclusions}

PM J13420-3415 largely owes its halo white dwarf status to its large
component of velocity perpendicular to the Galactic plane ($W>220$km
s$^{-1}$). This should be contrasted to the stars in the \citet{O01}
sample, which have $W<$100km s$^{-1}$ \citep{SRHKOB04}. Incidentally,
what distinguishes PM J13420-3415 from the \citet{O01} white dwarfs is
that it was discovered in a relatively low Galactic latitude field
($b=+27$), and its $W$ is largely reflected in its large proper
motion. In contrast, the \citet{O01} white dwarfs were found near the
south galactic cap, and a determination of their $W$ velocity is
entirely dependent on radial velocity measurements (which cannot be
obtained for many of them since their spectrum shows no atomic
line). It is clear that much would be gained from a survey of high
proper white dwarfs at low Galactic latitudes. Proper motion selection
would favor stars with large components of $W$, and determination of
halo membership would not be as critically dependent on the
determination of radial velocities. The identification of a large
sample of white dwarfs on near polar orbits could lift much of the
confusion that now exists between the thick disk and halo populations.

But while PM J13420-3415 has kinematics so clearly consistent with
a halo membership, its spectral energy distribution raises questions
about the origin its progenitor. One would expect a {\it bona fide}
halo white dwarf to be old, which generally means a much cooler
$T_{eff}$ than exhibited by PM J13420-3415. The luminosity function of
halo white dwarfs peaks at much fainter absolute magnitudes
\citep{IGHMT_98}, with a 10Gyr population peaking at $M_V\simeq16.5$,
and a 14Gyr population peaking at $M_V\simeq19$. With $M_V\approx15$,
PM J13420-3415 does not stand out as a typical remnant from early
generations of star.

It has been shown that relatively young, high-velocity white dwarfs
could originate in the thin disc, from which they can be ejected as
companions of stars undergoing catastrophic disruption. A Type Ia
supernova channel has been investigated by \citep{H_03}. However, the
resulting population of ejected white dwarfs is expected to have very
few objects with transverse velocities in excess of 200 km s$^{-1}$,
which would make PM J13420-3415 an exceptional case. Likewise, most
white dwarfs ejected following a Type II supernova events would have
motions generally consistent with a bloated thick disc \citep{DKR02}.

Alternatively, PM J13420-3415 could have been accreted into the Galaxy
in a merger event. The polar orbit displayed by PM J13420-3415 would,
for example, be consistent with the star having been accreted from the
Saggitarius dwarf Galaxy \citep{JMSRK99}. Should this be the case, PM
J13420-3415 could be associated with a local streamer. This hypothesis
would be corroborated if a whole group of stars were to be found in
the Solar Neighborhood with UVW velocities similar to PM J13420-3415.

It must however be pointed out that a $T_{eff}\approx5500$K white
dwarf does not preclude an old progenitor. PM J13420-3415 could be the
remnant of a moderately massive, 10-14Gyr old star that turned into a
white dwarf only $\sim$2Gyr ago. In that case, however, models predict
that the white dwarf should have a mass between $0.45M_{\odot}$ and
$0.50M_{\odot}$ \citet{FBB_01}. This could be tested through accurate
parallax measurements, which would provide a direct determination of
the absolute magnitude of PM J13420-3415, and hence its current
mass\footnote{Hypothetically, a massive white dwarf could also be
  compatible with old (MS+WD) age if it is the product of a merger of
  two low-mass degenerates; thanks to the referee (S. Vennes) for
  pointing this out.}. Indeed, mass determination now appears to be
critical in determining the true origin of white dwarfs with halo-like
kinematics \citep{BRHLCLD05}. A more precise distance combined with an
accurate estimate of the mass of PM J13420-3415 will thus be the key
in determining the true origin of the star. The results may have
significant implications about the white dwarf population in the
Galactic halo.

\acknowledgments

SL acknowledges support from Hilary Lipsitz, and from the American
Museum of Natural History.


\begin{thebibliography}{}

\bibitem[e.g. Bergeron(1993)]{B03}
Bergeron, P. 2003, \apj, 586, 201

\bibitem[Bergeron, Wesemael, \& Beauchamp(1995)]{BWB_95}
Bergeron, P., Wesemael, F., \& Beauchamp, A. 1995, \pasp, 107, 1047

\bibitem[Bergeron, Ruiz, \& Leggett(1997)]{BRL_97}
Bergeron, P., Ruiz, M. T., \& Leggett, S. K. 1997, \apjs, 108, 339

\bibitem[Bergeron {\it et al.}(2005)]{BRHLCLD05}
Bergeron, P., {\it et al.} 2005, \apj, 625, 838

\bibitem[Cutri {\it et al.}(2003)]{Cetal03}
Cutri R.M., et al. (2003) University of Massachusetts and Infrared
Processing and Analysis Center (IPAC/California Institute of
Technology)

\bibitem[Dauphole \& Colin(1995)]{DC95}
Dauphole, B., \& Colin, J. 1995, \aap, 300, 117

\bibitem[Davies, King, \& Ritter(2002)]{DKR02}
Davies, M. B., King, A.,\& Ritter, H. 2002, \mnras, 333, 463

\bibitem[Fontaine, Brassard, \& Bergeron(2001)(1995)]{FBB_01}
Fontaine, G., Brassard, P., \& Bergeron, P. 2001, \pasp, 113, 409 

\bibitem[Gates {\it et al.}(2004)]{Getal_04}
Gates, E., {\it et al.} 2004, \apj, 612, L129

\bibitem[Hansen(2003)]{H_03}
Hansen, B. M. S. 2003, \apj, 582, 915

\bibitem[Hansen \& Liebert(2003)]{HL03}
Hansen, B. M. S., \& Liebert, J. 2003, \araa, 41, 465

\bibitem[Ibata {\it et al.}(2000)]{IIBSG00}
Ibata, R., {\it et al.} 2000, \apj, 532, L41

\bibitem[Isern {\it et al.}(2003)]{IGHMT_98}
Isern, J., {\it et al.} 2003, \apj, 503, 239

\bibitem[Johnston {\it et al.}(1999)]{JMSRK99}
Johnston, K. V.,{\it et al.} 1999, \aj, 118, 1719

\bibitem[Koester(2003)]{K03}
Koester, D. 2003, \aapr, 11, 33

\bibitem[Latham {\it et al.}(2002)]{LSTDMCLM_02}
Latham, D. W., {\it et al.} 2002, \aj, 124, 1144

\bibitem[Liebert, Dahn, \& Monet(1989)]{LDM89}
Liebert, J., Dahn, C. C., \& Monet, D. G. 1989, in IAU Colloq. 114,
White Dwarfs, ed. G. Wegner (Berlin: Springer), 15

\bibitem[L\'epine, \& Shara(2005)]{LS05}
L\'epine, S., \& Shara, M. M. 2005, \aj, 129, 1483

\bibitem[L\'epine(2005a)]{L05a}
L\'epine, S. 2005, \aj, {\it in press}

\bibitem[L\'epine(2005a)]{L05b}
L\'epine, S. 2005, {\it in preparation}

\bibitem[Luyten (1979a)]{L79}
Luyten W. J. 1979, LHS Catalogue: a catalogue of stars
with proper motions exceeding 0.5" annually, University of Minnesota,
Minneapolis

\bibitem[Massey \& Gronwall(1990)]{MG90}
Massey, P., \& Gronwall, C. 1990, \apj, 358, 344

\bibitem[Monet {\it et al.}(2003)]{Metal03}
Monet D. G., et al. 2003, \aj, 125, 984

\bibitem[Oppenheimer {\it et al.}(2001)]{O01}
Oppenheimer, B. R., {\it et al.} 2001, Science, 292, 698

\bibitem[Reid(1996)]{R96}
Reid, I. N. 1996, \aj 111, 2000

\bibitem[Salim {\it et al.}(2004)]{SRHKOB04}
Salim, S., {\it et al.} 2004, \apj, 601, 1075

\bibitem[Scholz {\it et al.}(2002)]{SSAII_02}
Scholz, R.-D., {\it et al.} 2002, \apj, 565, 539

\bibitem[e.g. Spagna {\it et al.}(2005)]{SCLB05}
Spagna, A., {\it et al.} 2005, \aap, 428, 451


\end{thebibliography}
\end{document}